\documentclass[
reprint,
amsmath, 
amssymb, 
aps, 
superscriptaddress, 
pra]{revtex4-1}

\usepackage{graphicx}
\usepackage{dcolumn}
\usepackage{bm}
\usepackage{braket} 
\usepackage{hyperref}
\usepackage{subfigure} 
\usepackage{nccmath}
\usepackage{amsthm} 
\usepackage{mathrsfs} 
\usepackage{xfrac} 
\usepackage{pifont}
\usepackage{adjustbox}
\usepackage{bbm} 
\usepackage[utf8]{inputenc} 
\usepackage[normalem]{ulem}
\usepackage[usenames,dvipsnames,svgnames,table]{xcolor}
\usepackage[sort&compress]{natbib}
\usepackage{color}
\usepackage{longtable}
 
\usepackage{commath}
\usepackage{titlesec}

\begin{document}

\title{Loss-of-entanglement prediction of a controlled-{\footnotesize{PHASE}} gate in the framework of steepest-entropy-ascent quantum thermodynamics}

\author{J. A. Monta\~nez-Barrera}
 	\altaffiliation{ja.montanezbarrera@ugto.mx}
	\affiliation{Department of Mechanical Engineering, Universidad de Guanajuato, Salamanca, GTO 36885, Mexico}

\author{Cesar E. Damian-Ascencio}
 	\altaffiliation{cesar.damian@ugto.mx}
	\affiliation{Department of Mechanical Engineering, Universidad de Guanajuato, Salamanca, GTO 36885, Mexico}
	
\author{Michael R. von Spakovsky}
	\altaffiliation{vonspako@vt.edu}
	\affiliation{Department of Mechanical Engineering, Virginia Tech, Blacksburg, VA 24061, USA}

\author{Sergio Cano-Andrade}
	\altaffiliation{Corresponding author: sergio.cano@ugto.mx}
	\affiliation{Department of Mechanical Engineering, Universidad de Guanajuato, Salamanca, GTO 36885, Mexico}

\begin{abstract}
As has been shown elsewhere, a reasonable model of the loss of entanglement or correlation that occurs in quantum computations is one which assumes that they can effectively be predicted by a framework that presupposes the presence of irreversibilities internal to the system. It is based on the steepest-entropy-ascent principle and is used here to reproduce the behavior of a controlled-{\footnotesize{PHASE}} gate in good agreement with experimental data. The results show that the loss of entanglement predicted is related to the irreversibilities in a nontrivial way, providing a possible alternative approach that warrants exploration to that conventionally used to predict the loss of entanglement. The results provide a means for understanding this loss in quantum protocols from a nonequilibrium thermodynamic standpoint. This framework permits the development of strategies for extending either the maximum fidelity of the computation or the entanglement time.

\end{abstract}

\maketitle

\section{\label{int}INTRODUCTION}

Faster speeds and the ability to solve hitherto unsolvable problems are the motivations driving research on how to build quantum computers. The use of information theory in quantum systems goes back more than 30 years \cite{landauer1961irreversibility,zeh1970interpretation,kak1976uncertainty,Bennett1982,kak1984information,Feynman1982}. It has had two principal objectives: the characterization of information in a quantum system \cite{kak1976uncertainty,Bennett1982,kak1984information}, and the use of a quantum system to simulate a useful computation \cite{Feynman1982,deutsch1989quantum}. Current approaches to quantum computation exploit the phenomena of entanglement and superposition to create a paradigm that is more powerful than that of classical computing for certain types of problems \cite{shor1994algorithms,shor1999polynomial,lee2011entangling,tosi2017silicon}. 
However, one of the main problems in quantum computation is the loss of entanglement or correlation, which can take place at different timescales depending on the dynamics of the system \cite{Fischer2009}. Thus, successful implementation of a quantum computer requires the control of this loss.

Different experimental strategies for suppressing the loss of entanglement have been proposed and implemented over the last couple of decades. In Ref. \cite{Kuhlmann2013}, ultraclean and nuclear-spin free materials are used to reduce charge and spin noise, while in Ref. \cite{Viola1999} dynamical decoupling through echolike sequences is proposed to protect the information storage in the qubit from environmental fluctuations. Another proposed approach is to use dynamical decoupling to reduce the error using real-time feedback \cite{Biercuk2009}. Barthel \emph{et al}. \cite{Barthel2010} suggest that the loss-of-entanglement recovery in singlet-triplet qubits can be achieved by using dynamical decoupling. Still another approach is to control the nuclear spin bath conditions by suppressing the qubit dephasing with a nuclear state preparation \cite{Reilly2008}. Recent research has also explored how a quantum computer can be controlled using electrostatically coupled quantum dots \cite{Loss1997,shulman2012demonstration,ghirri2014quantum,vandersypen2017interfacing,Nichol2017}. This has resulted in new ways of constructing and manipulating qubits and has increased our understanding of the mechanisms involved in the evolution of quantum information systems.

In addition to experimental techniques, dealing with the loss of entanglement requires understanding how it is generated and how it is related to the system's state evolution. Thus, theoretical approaches are also needed to gain a clearer understanding of this phenomenon. A common model used to describe this loss is to assume that the system of interest interacts with a thermal bath (reservoir or environment) and that the loss of entanglement is a consequence of the weak interactions that exist between the system and the bath. Using the Markov approximation, the state evolution of the system interacting with the bath is modeled using a linear Markovian quantum master equation of the Kossakowski-Lindblad-Gorini-Sudarshan type \cite{Lindblad1976,Yuan2007Markovian,Nakatani2010Quantum}. This methodology is used in Ref. \cite{Rajagopal2001} to predict the loss of entanglement during the evolution of a pair of coupled quantum dissipative oscillators, and in Ref. \cite{Bertlmann2006}, this methodology is used to predict the loss of entanglement on initially entangled qubits. Another model for predicting this loss is Milburn's model, which assumes that the system evolves as a sequence of random unitary transformations for short periods of time \cite{Milburn1991}. This model is used in Ref. \cite{Kimm2002} to study the evolution of a two-qubit quantum swap gate.

Other recent models of decoherence due to hyperfine interactions have shown some success in modeling the dephasing and the Hahn echo revival phenomenon occurring in quantum systems in the presence of low-magnetic fields ($B_\mathrm{ext} < 3$ T). For example in Ref. \cite{Bluhm2011}, the nuclear-spin-induced decoherence is explained with a semiclassical model which accounts for the dynamics of the electron and nuclear spins to predict the effects due to nuclear hyperfine interactions. This effect is dominant in the presence of magnetic fields under 400 mT. The model does a good job of predicting the Hahn echo revivals and is based on the model of Ref. \cite{Cywinski2009}, which is a decoherence model for a spin qubit interacting with a nuclear spin bath that shows some qualitative and quantitative success in predicting the effects of the hyperfine interactions caused by nuclear spins in the presence of low-magnetic fields and short timescales. They propose that for these conditions, the differences in the Zeeman energies between interacting nuclei are the main source of decoherence. In Ref. \cite{Bluhm2011}, a pair-correlation method for electron-nuclear-spin dynamics is presented as an explanation of the electron spin decoherence due to hyperfine interactions. It offers an explanation of quantum decoherence in terms of entanglement with a thermal bath.

Still other models describe the loss of entanglement without dissipation where the only effect is that of elementary quantum mechanics and equilibrium statistical mechanics \cite{Ford2001,Fortin2014}. Yet another approach is that based on the framework of steepest-entropy-ascent quantum thermodynamics (SEAQT)  \cite{beretta1984quantum,Beretta1985,beretta2010maximum}, which has been shown to be a reasonable approach for describing the phenomenon of the loss of entanglement as well as the coherence present during the evolution of quantum composite systems \cite{Cano2015} such as those used in quantum gate operations.

It is this last approach, the SEAQT framework  \cite{beretta1984quantum,Beretta1985,Beretta2014,Beretta2009,beretta2010maximum,Beretta2006,vonSpakovsky2014,Cano2015,Li2016,Li2016a,Li2018,Li2018a,smith2016comparing}, which is used here to model the two-qubit operation of Shulman \emph{et al.} \cite{shulman2012demonstration} and which shows good agreement with the experimental results given in Ref. \cite{shulman2012demonstration}. In doing so, it provides an understanding of the loss of entanglement in the quantum protocols from a quantum thermodynamic standpoint \cite{Cano2015}.

The rest of the paper is organized as follows. Section \ref{Syst} provides a description of the controlled-{\footnotesize{PHASE}}  ({\footnotesize{CPHASE}} ) gate of Shulman \emph{et al.}   \cite{shulman2012demonstration}, while Sec. \ref{MatMod} provides a description of the SEAQT model and some measures of entanglement and the closeness of two states. Section \ref{Result} then presents the results and discussion. Section \ref{RelDisDecay} presents a presentation of the relationship between the dissipative time and the decay rate for the particular {\footnotesize{CPHASE}} gate studied. Finally, Sec.  \ref{Conc} provides some conclusions.

\section{\label{Syst}CPHASE GATE}

\begin{figure}[]
\includegraphics[width=8.8cm]{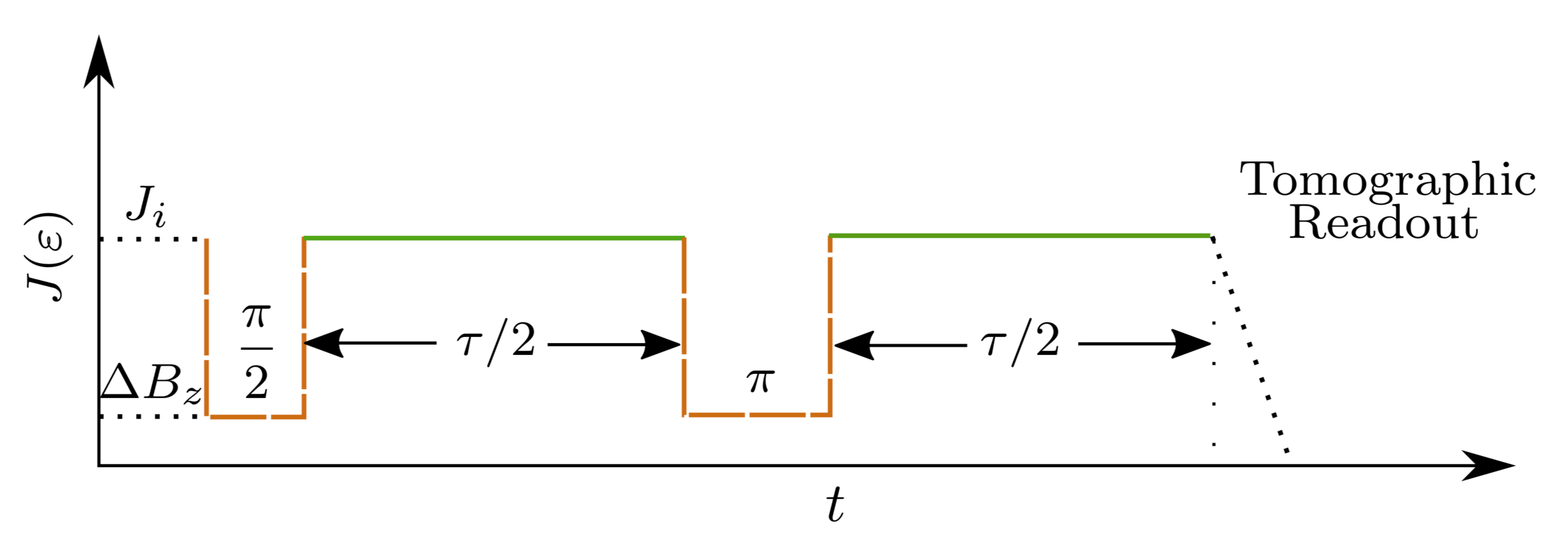}
\caption{\label{protocol}Controlled-{\scriptsize{PHASE}} quantum gate protocol by Shulman \emph{et al.} \cite{shulman2012demonstration}.}
\end{figure}

The quantum protocol is investigated by Shulman \emph{et al.} \cite{shulman2012demonstration}. The experiments are based on the implementation of a singlet-triplet ($S-T_0$) qubit via the confinement of two electrons to a double quantum dot (QD) in a two-dimensional electron gas situated below a GaAs-AlGaAs heterostructure surface \cite{shulman2012demonstration}. The {\footnotesize{CPHASE}} gate protocol is shown in Fig. \ref{protocol} and is as follows. The qubits are initialized in the $\ket{S}$ state and then a $\pi/2$ rotation around the $x$ axis is applied using a magnetic field gradient of $\Delta B_{z_i}/ 2 \pi = 30$ MHz between the electrons with $J(\varepsilon)_i =0$. After this, the qubits are allowed to evolve during a time  $\tau/2$ with $J_1 / 2 \pi \approx 280$ MHz and $J_2 / 2 \pi \approx 320$ MHz  $\gg \Delta B_z$, causing the qubits to rotate around the $z$ axis. Next, with $J_i =0$, a $\pi$ rotation is applied around the $x$ axis, which causes the qubits to decouple from the environment, while still maintaining the coupling with each other. Subsequently, the qubits are allowed to evolve during a time $\tau/2$ after which measurements are performed on the state of the system \cite {shulman2012demonstration}. Several experiments are performed in terms of the difference in energy, $\varepsilon$, between the levels of the QDs, and its implications on the outcome are measured in terms of the fidelity and concurrence, both measures of the coherence.

The Hamiltonian of this {\footnotesize{CPHASE}} gate is represented by

\begin{eqnarray}\label{Eq7}
H_\mathrm{CPG} = \frac{\hbar}{2} [ J_1 (\sigma_z \otimes I ) +J_2 ( I \otimes \sigma_z )   \nonumber \\
+ \frac{J_{12}}{2} ( \sigma_z - I ) \otimes ( \sigma_z - I ) \\ 
+ \Delta B_{z,1} ( \sigma_x \otimes I ) + \Delta B_{z,2} ( I \otimes \sigma_x ) ] \nonumber \, ,
\end{eqnarray}
where $\sigma_{x}$, $\sigma_{y}$, and $\sigma_{z}$ are the standard Pauli matrices; $J_i$ $(i=1,2)$ is the exchange spliting; $\Delta B_{z,i}$ $(i=1,2)$ is the magnetic field gradient; and $J_{12}$ is the two qubit coupling which is proportional to the product of the dipole moments for each qubit. The first two terms in Eq. (\ref{Eq7}) produce the Zeemann effect, and, as is pointed out in Ref. \cite{shulman2012demonstration}, the third term of the Hamiltonian is responsible for the coupling of both qubits that produces the entangled state. The last two terms give rise to the dynamical decoupling. In the rotational frame, the Hamiltonian is written as
\begin{equation}
H_\mathrm{CPG, rot} = \hbar \frac{J_{12}}{4}\left( \sigma_z - I \right)\otimes \left( \sigma_z - I \right), 
\end{equation}
which is the relevant term in the Hamiltonian given by Eq. (\ref{Eq7}) responsible for the entanglement.

\section{\label{MatMod}MATHEMATICAL MODEL}

\subsection{The SEAQT framework}

In the SEAQT framework, the dynamics of the density or ``state'' operator, $\rho$, of a quantum system is governed by both a symplectic (unitary) and a relaxation (nonunitary) term that, respectively, capture both the reversible and irreversible dynamics of state evolution. The former is the so-called von Neumann term of quantum mechanics, while the latter is based on the principle of steepest entropy ascent (SEA), which at every instant of time drives the state evolution in the direction of maximal entropy increase. Note that the view of physical reality assumed by this framework is one in which the nonlinear dynamics of state relaxation are intrinsic to the system and not a consequence of interactions with an environment. This contrasts with the standard open quantum system model presented in the next section. The SEAQT equation of motion for a general quantum system \cite{Beretta1985} consisting of two-qubits is written as
\begin{equation}\label{Eq:Beretta}
\frac{d\rho}{dt} = -\frac{i}{\hbar}[H,\rho] - \left(  \frac{1}{\tau_{D_1}} D_1\otimes \rho_2 +  \frac{1}{\tau_{D_2}} \rho_1\otimes D_2\right),
\end{equation}
where $H$ and $\rho$ are the Hamiltonian and the density operator for the composite system, the $\rho_J (J = 1, 2)$ are the density operators for qubits 1 and 2, the $\tau_{D_J} (J = 1, 2)$ are internal-relaxation times that are positive constants or positive functionals of the $\rho_J$ , and the $D_J (J = 1, 2)$ are the dissipation (or relaxation) operators for each qubit. The latter are Hermitian and written as

\begin{equation}
D_J = \frac{1}{2} [ \sqrt{\rho_J} \tilde{D}_J + (\sqrt{\rho_J}\tilde{D}_J)^{\dagger} ],
\end{equation}
where the symbol $\dag$ signifies the adjoint and each $\tilde D_J$ for this two-qubit system is expressed as

\begin{equation} \label{EqtildeD_J}
\tilde{D}_J = \frac{
\begin{vmatrix}
\sqrt{\rho_J}(B\ln{\rho})^J & \sqrt{\rho_J}(I)^J & \sqrt{\rho_J}(H)^J \\
(I,B \ln{\rho})^J &(I,I)^J&(I,H)^J\\
(H,B \ln{\rho})^J&(H,I)^J&(H,H)^J\\
\end{vmatrix}
}{
\begin{vmatrix}
(I,I)^J & (I,H)^J\\
(H,I)^J & (H,H)^J\\
\end{vmatrix}
} .
\end{equation}
Here $(\cdot,\cdot)^J$ is the Hilbert-Schmidt inner product defined on Hilbert space $\mathcal{H}^J$ by $(F,G)^J=\textrm{Tr}_J[\rho_J\{(F)^J,(G)^J\}]$ with $J=A$, $B$, $(F)^A=\textrm{Tr}_B[(I_A\otimes\rho_B)F]$, and $(F)^B=\textrm{Tr}_A[(\rho_A\otimes I_B)F]$. In Eq. (\ref{EqtildeD_J}), $B$ is the projector onto the range of $\rho$, i.e., the idempotent operator that results from summing up all the eigenprojectors of $\rho$ belonging to its nonzero eigenvalues. For more details, the reader is referred to Refs. \cite{beretta1984quantum,beretta2010maximum}. 

The relaxation is directly related to the entropy via the standard von Neumann entropy given by
\begin{equation}
S = - k_B \mathrm{Tr} \left( \rho \ln \rho \right),
\label{Eq:Entropy}
\end{equation}
where $k_B$ is the Boltzmann's constant so that the rate of entropy generation is expressed as 
\begin{equation}
	\frac{dS}{dt} = - k_B \frac{d}{dt} \mathrm{Tr} \left( \rho \ln \rho \right),
\label{Eq:EntropyProd}
\end{equation}

\begin{equation}
\frac{dS}{dt}=-\frac{\left|
	\begin{array}{ccc}
	(B\ln\rho,B\ln\rho)^J & (B\ln\rho,I)^J & (B\ln\rho,H)^J\\
	(I,B\ln\rho)^J & (I,I)^J & (I,H)^J\\
	(H,B\ln\rho)^J & (H,I)^J & (H,H)^J \end{array}\right|}{\left|
	\begin{array}{cc}
	(I,I)^J & (I,H)^J\\
	(H,I)^J & (H,H)^J \end{array}\right|}.
\end{equation}

As indicated in Ref. \cite{shulman2012demonstration} (Supplementary Information), the evolution of the state of the  $S-T_0$ qubit system starts with a Bloch vector modulus of 0.95. This value is used here as the initial state for the evolution of the SEAQT equation of motion.

\subsection{The open quantum system model}

An alternative approach to the one proposed above is to use one of the equations of motion of the theory of open quantum systems \cite{Breuer2002}. In this approach, the equation of motion uses a reduced density operator that may be viewed as originating from the concept of typicality as discussed in Ref. \cite{vonSpakovsky2014}. For Markovian master equations of the Lindblad type \cite{Breuer2002,Lindblad1976},  the open quantum system model assumes that weak coupling, i.e., statistical perturbations (the so-called Born-Markov approximation), exists between the system and its surroundings. This approach, thus, relies on a partitioning between the primary system and the environment and assumes that the total state evolution of the composite system-environment is unitary and generated by a composite Hamiltonian. The evolution of the system state alone is then based on a reduced dynamics, which leads to the appearance of a so-called ``dissipative'' state evolution. Of course, as shown by Nakatani and Ogawa \cite{Nakatani2010Quantum}, a limitation of this approach is that the Born-Markov approximation for obtaining evolution equations cannot be used for composite systems in the strong-coupling regime.
	
The equation of motion of the Lindblad type used here is expressed as

\begin{equation}\label{eq8}
\frac{d \rho_s}{dt} = -\frac{i}{\hbar} [H,\rho_s] + \frac{\gamma}{2} \displaystyle\sum_j (  2L_j \rho_s L_j^\dagger - L_j^\dagger L_j \rho_s  -  \rho_s  L_j^\dagger  L_j),
\end{equation}
where $H$ is the Hamiltonian, $\rho_s$ is the reduced density operator of the system, the $L_j$ are the Lindblad operators which represent the coupling of the system with the environment, and $\gamma$ is the spontaneous decay rate. Note that the right hand side of Eq. (\ref{eq8}) is a linear superoperator operating on $\rho_s$ and represents a semi-group. This contrasts with the SEAQT equation of motion which represents a full group and for which the dynamics of its dissipation operator is nonlinear. Furthermore, even though the open quantum system model is based on the orthodox belief that unitary (linear) dynamics is foundational and that as a consequence the second law of thermodynamics emerges from quantum mechanics, this assumption cannot, as pointed out in Refs. \cite{Gheorghiu2001,Domokos1999,Czachor1998a,Czachor1998b,Czachor1999,Jordan1993}, rule out the possibility of a non-linear dynamics such as, for example, that seen in the SEAQT formulation since ``\emph{if the pure states happen to be attractors of a nonlinear evolution, then testing the unitary propagation of pure states alone cannot rule out a nonlinear propagation of mixtures}'' \cite{Gheorghiu2001}.    

Now, for the particular case studied here, the Lindblad operators are given by \cite{Saki2019}

\begin{equation}
L_1 =  I \otimes L_\mathrm{phase},
\end{equation} 

\begin{equation}
L_2 = L_\mathrm{phase} \otimes I,
\end{equation} 
where 
\begin{equation}
L_\mathrm{phase} = \sqrt{\lambda} \ \sigma_z
\end{equation} 
and  $\lambda$ is the parameter defining the strength of the phase damping.

\subsection{Measures of entanglement and closeness}

To verify the entanglement of the qubits, the concurrence defined as
\begin{equation}\label{concurrence}
C(\rho) = \mathrm{max} \{ 0, \lambda_4-\lambda_3-\lambda_2-\lambda_1 \}
\end{equation}
is used. Here the $\lambda_i$ are the eigenvalues, sorted from smallest to largest, of the matrix $R = \sqrt{\sqrt{\rho} \tilde \rho \sqrt{\rho}}$ where $\tilde \rho = (\sigma_y \otimes \sigma_y ) \rho^* ( \sigma_y \otimes \sigma_y )$ and $\rho^*$ is the complex conjugate of $\rho$. A positive value between 0 and 1 for $C(\rho)$ indicates that entanglement between the qubits is present, while a value of $C(\rho) \leq 0$ means that there is no entanglement.

To measure the closeness of the desired state, the Bells state fidelity is used. It is expressed as
\begin{equation} \label{eq:fidelity}
F = \langle \Phi_\mathrm{ent} | \rho | \Phi_\mathrm{ent} \rangle,
\end{equation}
where $| \Phi_\mathrm{ent} \rangle = \exp\left[ i \pi \left( I \otimes \sigma_y + \sigma_y \otimes I \right) /8  \right] | \Psi_- \rangle$ is a generalized Bell state, while $| \Psi_- \rangle = \frac{1}{\sqrt{2}} \left( | SS \rangle - | T_0 T_0 \rangle \right)$ is a Bell state. Here $\sigma_y$ is the $y$-Pauli operator, $I$ is the identity operator, and $\ket{S} = \frac{1}{\sqrt{2}}(\ket{\uparrow\downarrow} - \ket{\downarrow\uparrow})$ and $\ket{T_0} = \frac{1}{\sqrt{2}}(\ket{\uparrow\downarrow} + \ket{\downarrow\uparrow})$ are the basis states for the $S-T_0$ (singlet-triplet) qubits. For all non-entangled states, $F \leq 0.5$ \cite{shulman2012demonstration}. Thus, the generalized Bell state differs from the Bell state by a unitary transformation.

\section{\label{Result}RESULTS AND DISCUSSION}

\subsection{Concurrence and fidelity}

\begin{figure}[]
\includegraphics[width=9.5cm]{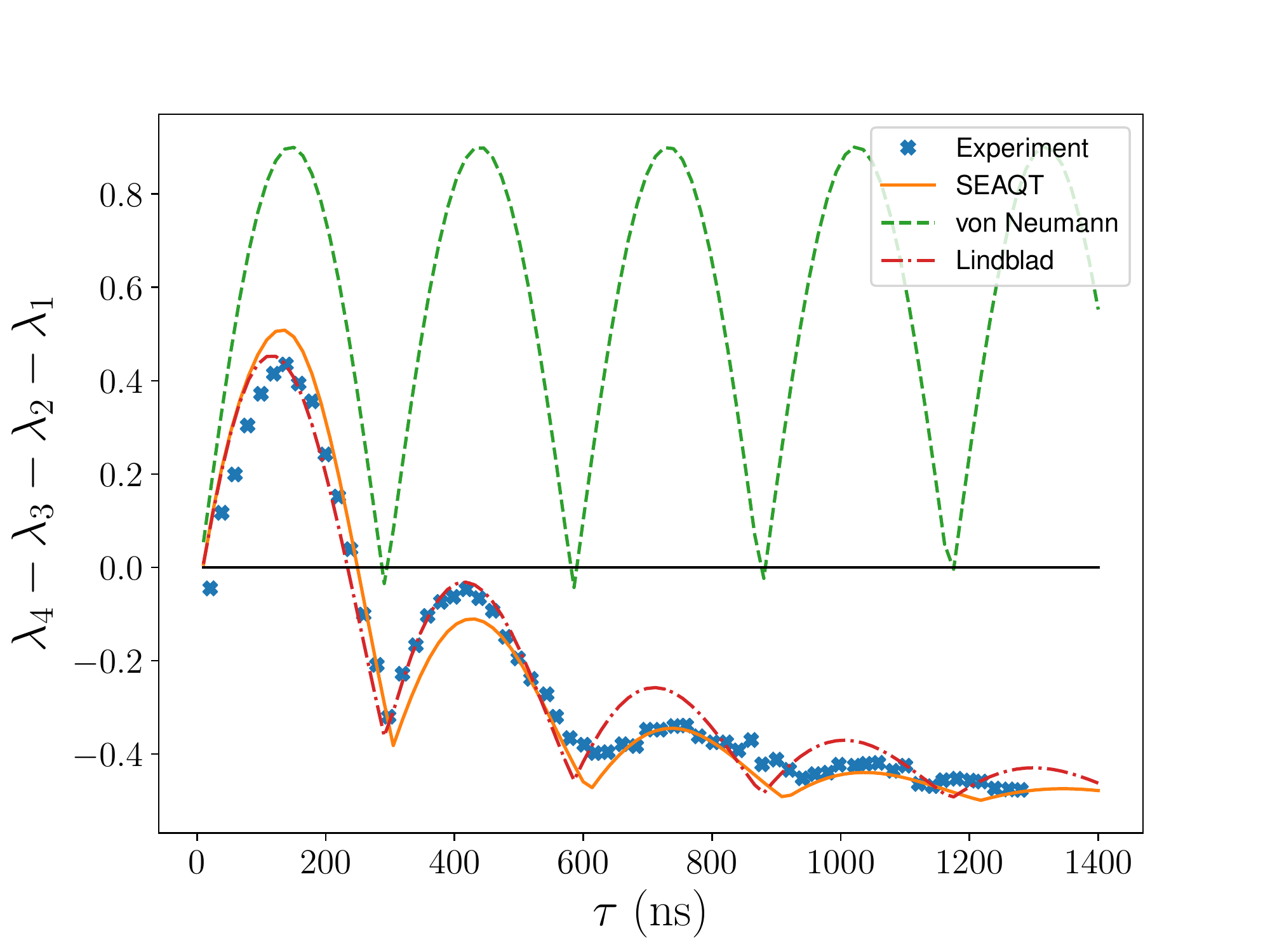} \\ (a)
\includegraphics[width=9.5cm]{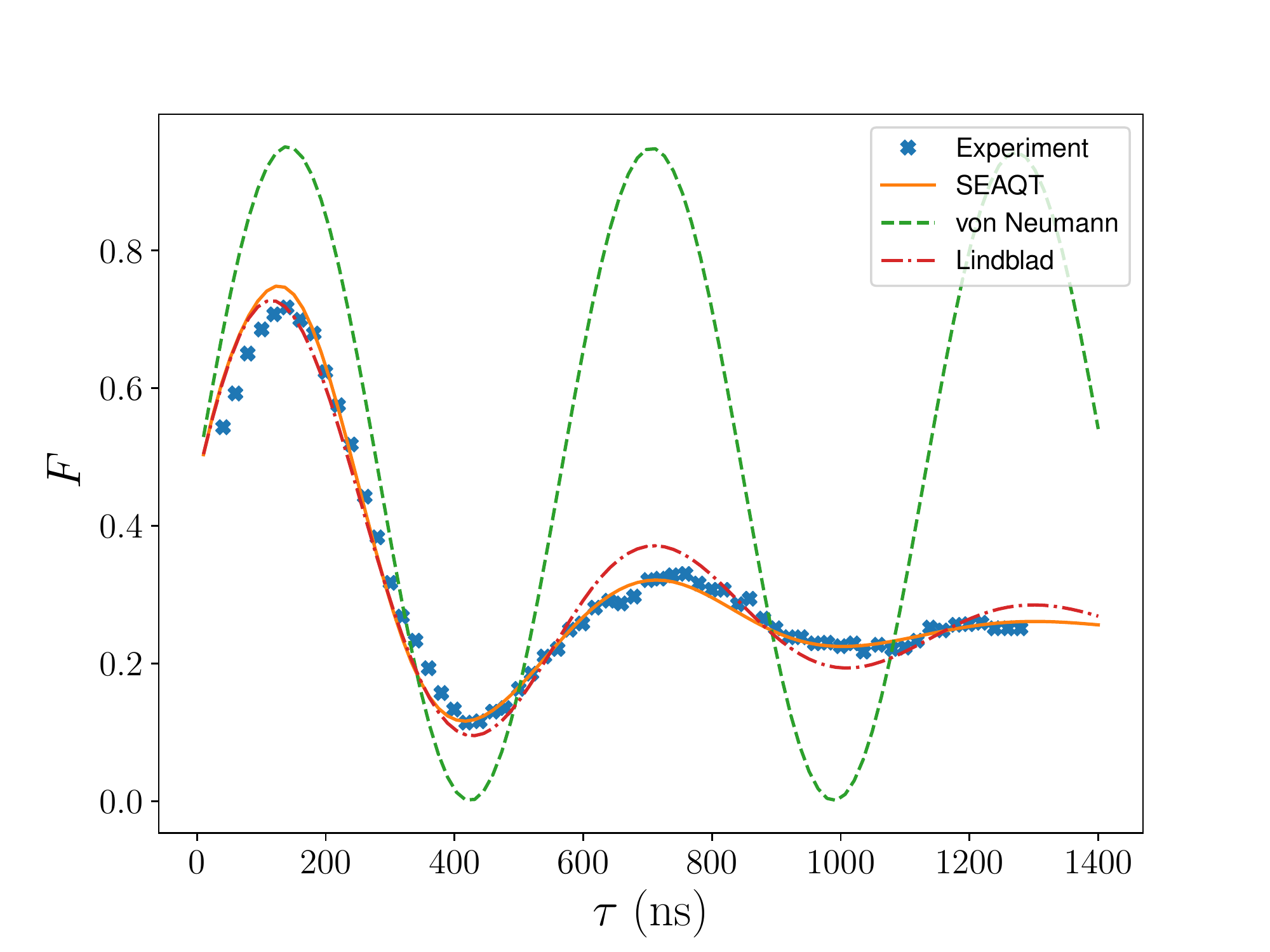} \\ (b)
\caption{\label{fig-2n} (a) Difference in the sorted eigenvalues of the $R$ matrix for $\delta{\varepsilon}=80 \, \mu$V of the final state density operator, $\rho(\tau)$, for different values of the {\scriptsize{CPHASE}} gate duration time, $\tau$. A positive value indicates entanglement. (b) The fidelity with which the final state density operator resembles that of the Bell state for different values of the duration time $\tau$.}
\end{figure}

Experimentally, it has been demonstrated that the concur- rence decreases with time and only keeps a positive value for short periods of time. Figure \ref{fig-2n} shows the experimental values reported in Ref. \cite{shulman2012demonstration} for the concurrence and fidelity of the final state as a function of the duration $\tau$, of the {\footnotesize{CPHASE}} gate operation. Also seen in this figure are the predictions obtained from the SEAQT formulation as well as those from the Lindblad and von Neumann equations. The measured data, i.e., each blue cross, appearing in Fig. \ref{fig-2n} is obtained in Ref. \cite {shulman2012demonstration} for a maximum entanglement time, $\tau_\mathrm{ent}$, corresponding to a difference in energy, $\delta{\varepsilon}$, of 80 $\mu$V between the levels of the left and right QDs. As seen, the experimental data decreases its amplitude with each successive oscillation. These results suggest that, as time passes, the entanglement between qubits is lost at a value of the {\footnotesize{CPHASE}} duration time, $\tau$, greater than about 240 ns. At this point the difference in the $R$ matrix eigenvalues shown in Eq. (\ref{concurrence}) becomes negative and stays negative.

Figure \ref{fig-2n}(a) shows the results for the concurrence. The green dashed curve is the result obtained here with the von Neumann equation, i.e., the symplectic part only of the SEAQT equation of motion. As seen, the results of the von Neumann equation never decrease, oscillating between zero and some maximum amplitude. The predictions of the Lindblad equation are shown by the red dashed-dotted curve. As can be seen, these predictions are in good agreement with experimental data for the first and second oscillations but start to deviate for subsequent oscillations (after about 600 ns). However, since only part of the first oscillation is positive, the Lindblad equation predicts the loss of entanglement between the qubits well. The SEAQT predictions are shown by the orange solid curve. These predictions are also in good agreement with the experimental data throughout the entire evolution. This suggests that both models resulting from very different views of physical reality are able to predict the concurrence and hence the entanglement in time.

The fidelity is used here to measure how far the final state is from the expected final state (i.e., in our case, the Bell state) for a given protocol. Figure  \ref{fig-2n}(b) presents a comparison between the final fidelity obtained with the experimental data of Ref. \cite{shulman2012demonstration} and those determined by the SEAQT equation of motion, the von Neumann equation of motion, and the Lindblad equation. Again, the predictions made with the von Neumann equation oscillate without any net loss of fidelity. In contrast, the Lindblad predictions follow the decrease seen in the experimental data, although again with some deviations in amplitude after the first oscillation. The SEAQT predictions also follow the experimental data very closely over the entire evolution but more closely than the Lindblad predictions. As before, this suggests that the SEAQT framework is a viable approach for predicting the loss of entanglement in time.

\begin{figure}[]
\includegraphics[height=10cm]{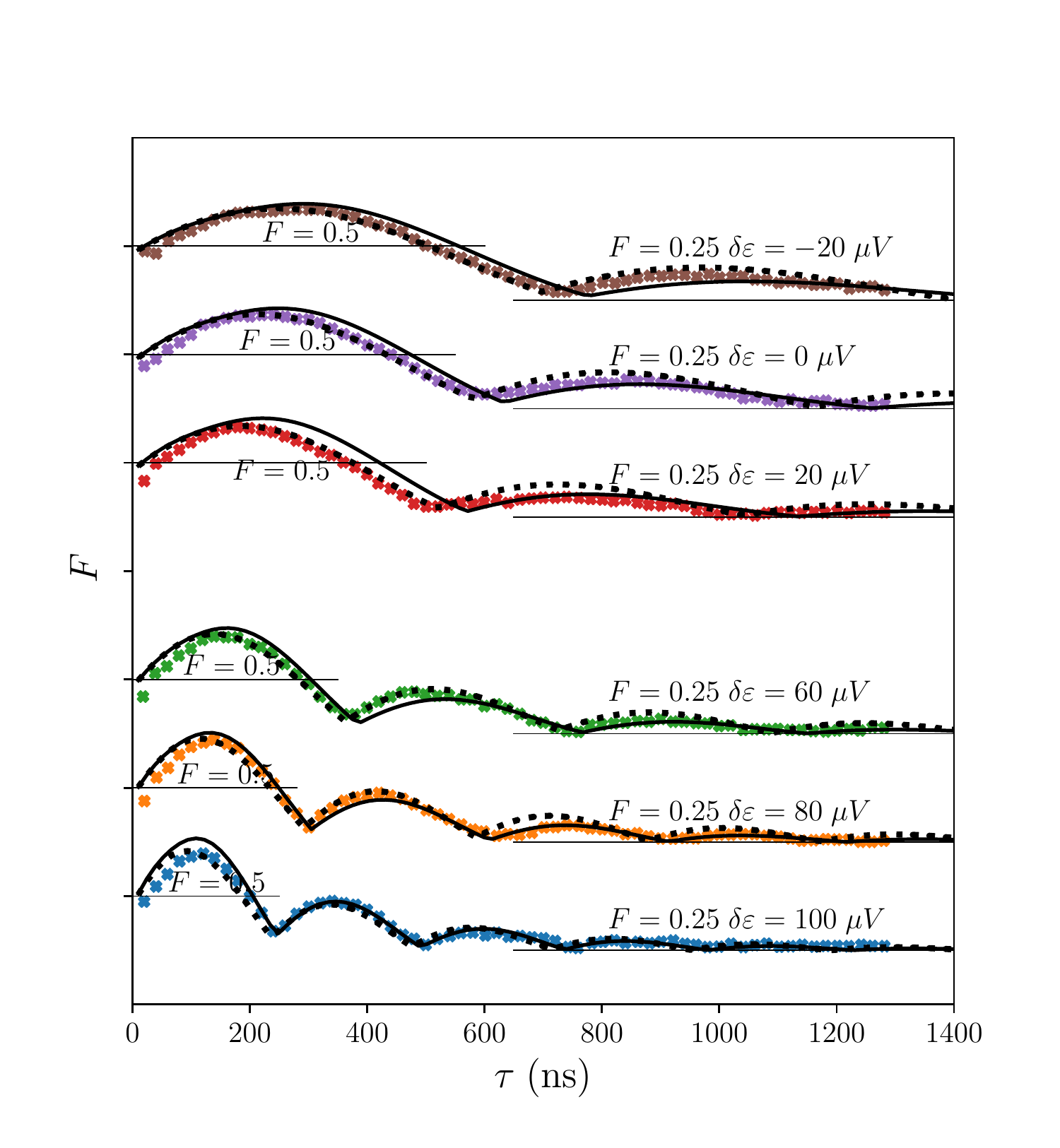}
\caption{\label{fig-3n} Fidelity for different values of $\delta\varepsilon$ as a function of the {\scriptsize{CPHASE}} duration time, $\tau$. The solid black line represents the SEAQT prediction and the black dotted line represents the Lindblad prediction for the different values of  $\delta\varepsilon$.}
\end{figure}

Figure \ref{fig-3n} shows a comparison between the results predicted by the SEAQT and Lindblad equations of motion for different values of $\delta\varepsilon$ and the experimental results. Here, both equations of motion, i.e., the SEAQT and Lindblad, predict that the variation in $\delta{\varepsilon}$ to negative values increases the maximum entanglement time $\tau_\mathrm{ent}$, without, however, an improvement in the maximal value of the fidelity obtained, which is consistent with what is observed experimentally.

\subsection{Thermodynamics of the {\footnotesize{CPHASE}} protocol}

\begin{figure}[]
	\includegraphics[width=9.5cm]{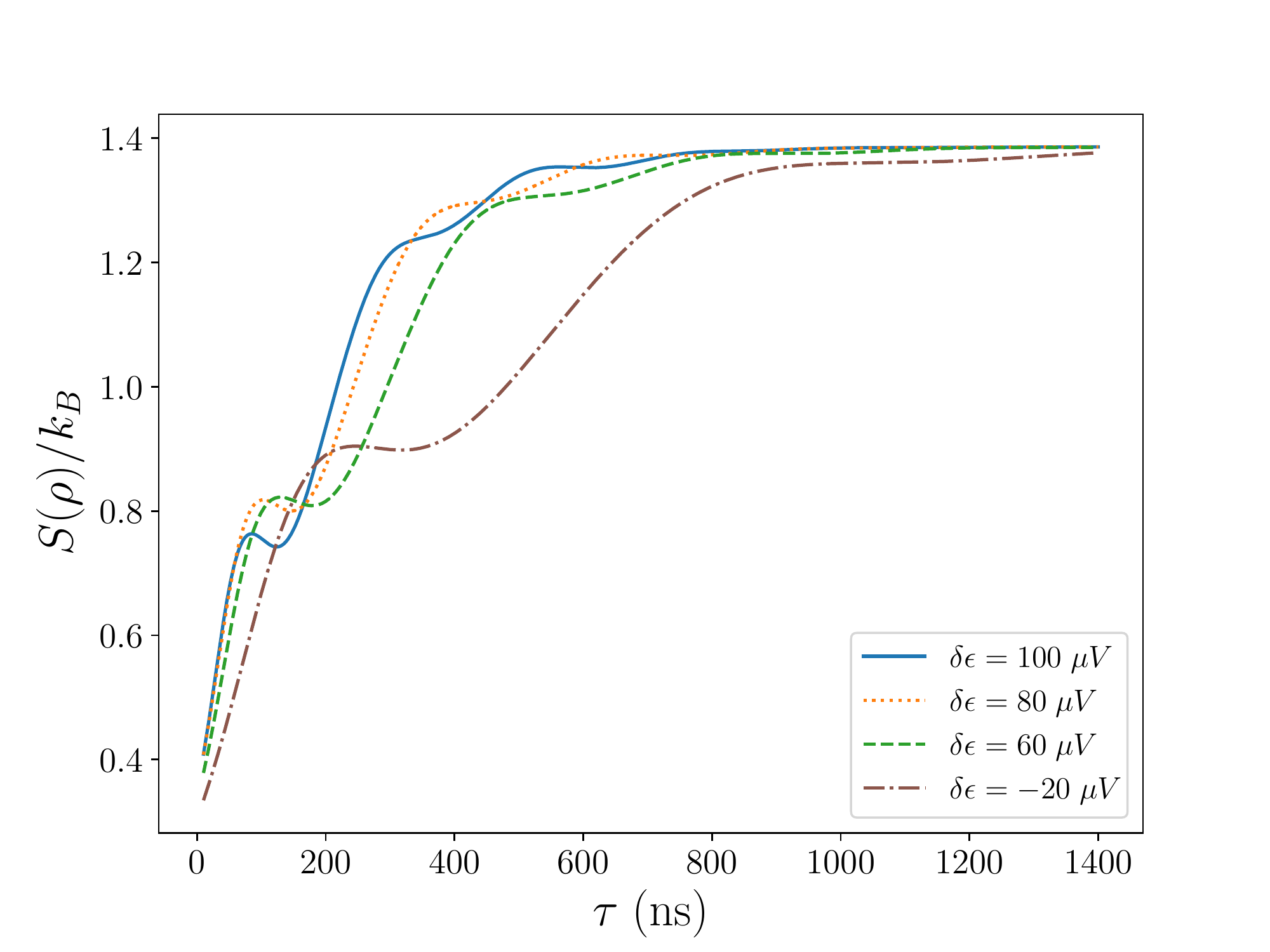} \\ (a) 
	\includegraphics[width=9.5cm]{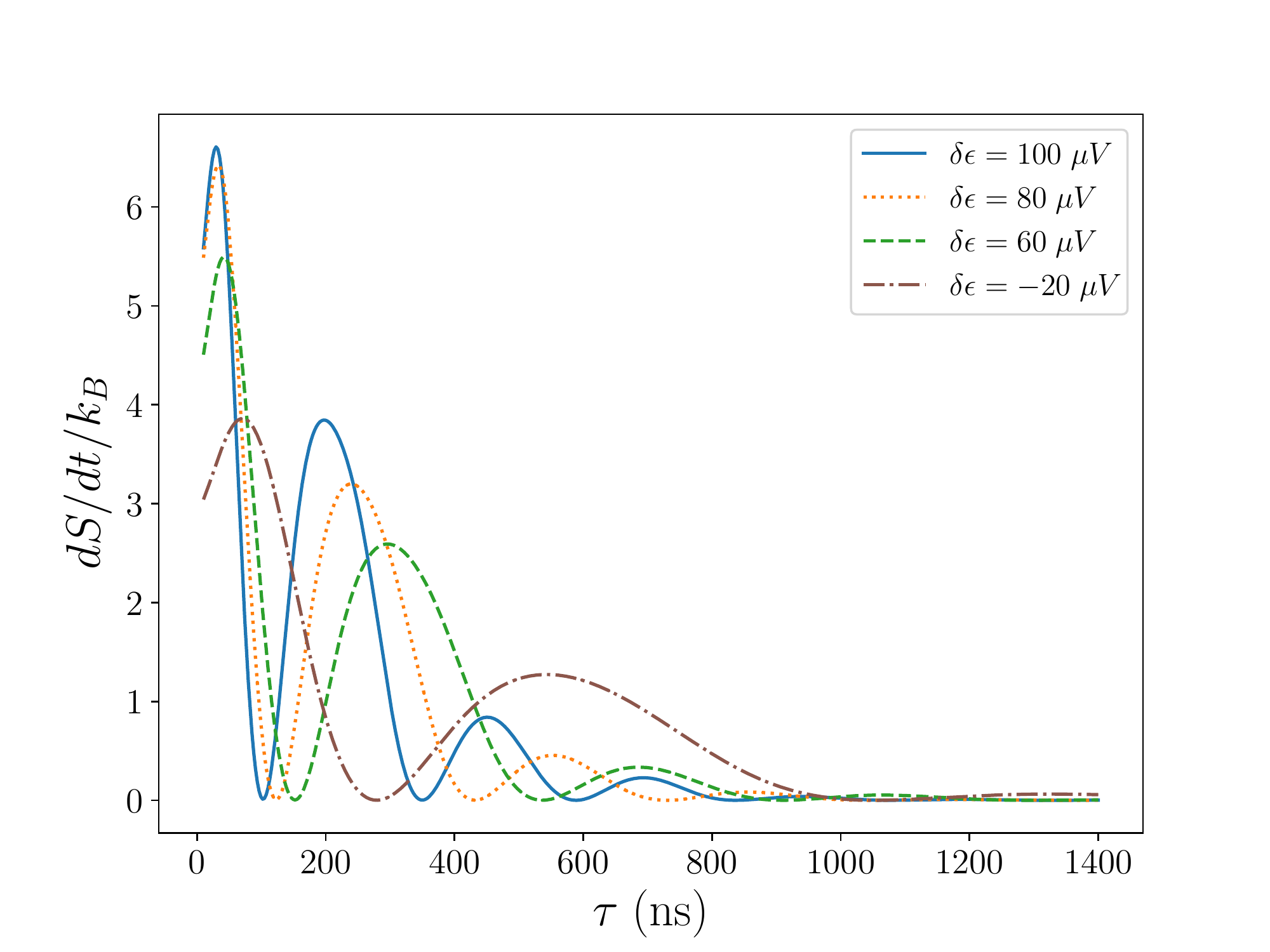} \\ (b) 
	\caption{\label{fig-4n}Final (a) entropy and (b) entropy generation rate (or equivalently the rate of entropy change) predicted by the SEAQT equation of motion for different values of $\delta\varepsilon$ and the duration time $\tau$. This quantities are plotted in dimensionless form as $S/k_B$ and $dS/dt / k_B$.}
\end{figure}

\begin{figure}[]
	\includegraphics[height=3cm]{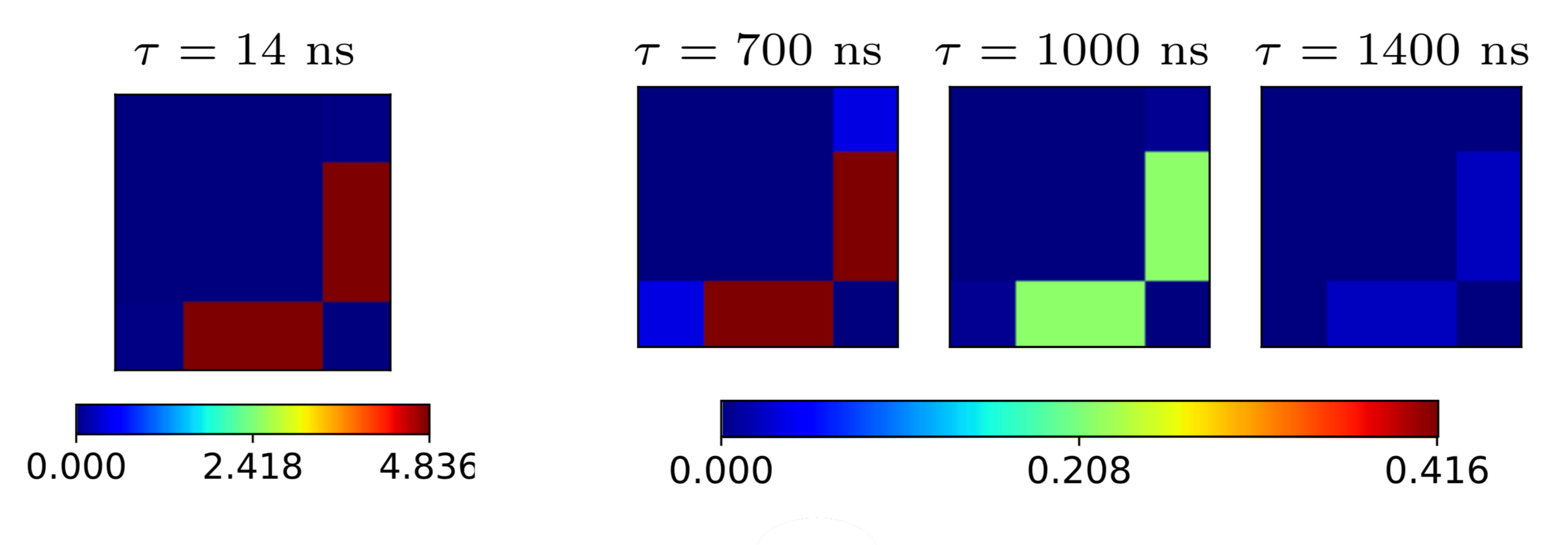} \\ a) 
	\\
	\includegraphics[height=2.7cm]{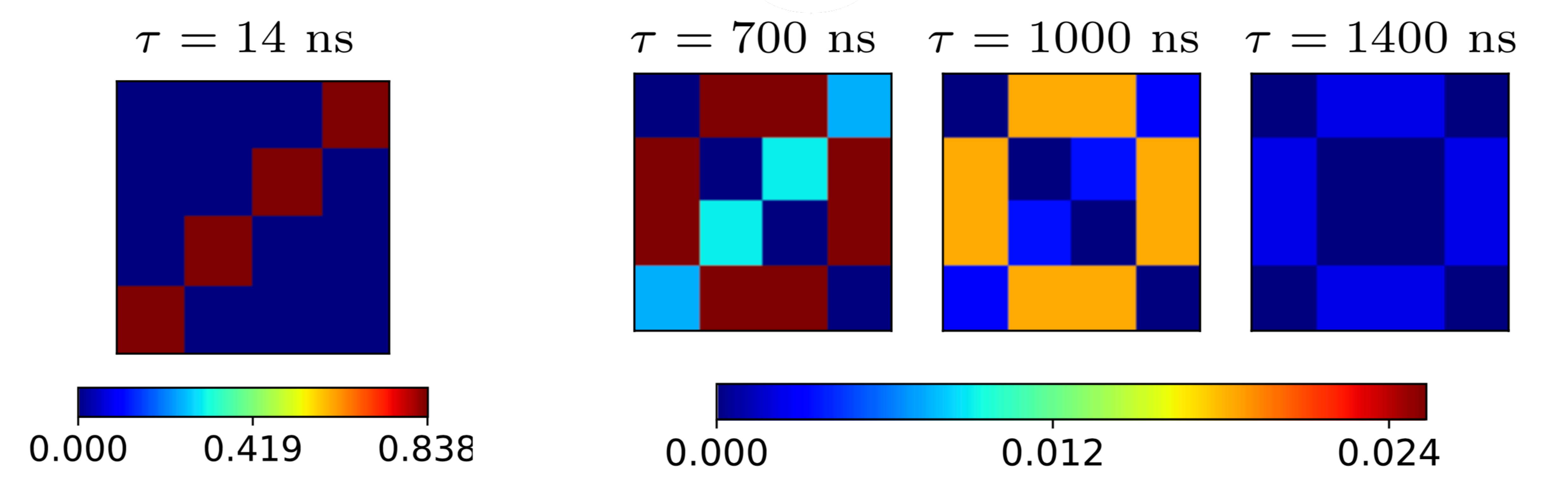} \\ b) 
	\caption{\label{SEAQTterms}Final state of the {\scriptsize{CPHASE}} protocol in terms of the (a) von Neumann and (b) dissipative terms of the SEAQT equation of motion for $\tau = 14$ ns (left panels), $\tau = 700$ ns (middle left panels), $\tau =1000$ ns (middle right panels), and $\tau =1400$ ns (right panels). A value of $\delta\varepsilon = 80$  $\mu$V is considered. These values are normalized using a time of 1 $\mu$s.}
\end{figure}

In Fig. \ref{fig-4n}, the final entropy and entropy generation rate (equivalent in this case to the rate of entropy change) for the SEAQT framework of each {\footnotesize{CPHASE}} gate execution as a function of the {\footnotesize{CPHASE}} gate duration time $\tau$ are presented for four different values of $\delta{\varepsilon}$. The entropy and entropy generation are central to the SEAQT equation of motion, which determines the unique nonequilibrium thermodynamic path of state evolution based on maximal entropy generation. The direction taken is either towards a state of stable equilibrium for which both the symplectic and dissipative terms of the equation of motion go to zero or towards a stationary state in which only the latter term vanishes. As seen in Fig. \ref{fig-4n}(a), the final entropy values are not monotonically increasing, while in Fig. \ref{fig-4n}(b) it is observed that, for duration times greater than 1000 ns, the entropy generation rate goes to zero indicating that a state of stable equilibrium is reached for each value of $\delta{\varepsilon}$. This is also observed in Figure \ref{SEAQTterms} were the $4\times 4$ matrices corresponding to the von Neumann [Fig. \ref{SEAQTterms}(a)] and dissipative [Fig. \ref{SEAQTterms}(b)] terms of the SEAQT equation of motion are shown for the final state of four different values of $\tau$ with $\delta\varepsilon = 80$  $\mu$V. It is observed that for the value of $\tau = 1400$ ns, the von Neumann and dissipative terms are effectively zero and, thus, the final state is that of stable equilibrium. For the other cases, i.e., $\tau = $ 400 ns, 700 ns, and 1000 ns, neither the dissipative nor the von Neumann term is zero and, thus, continued evolution of these final states is possible.

In looking at both Figs.  \ref{fig-4n}(b) and \ref{fig-3n} for the case of, for example, $\delta{\varepsilon}=100 \ \mu$V, the points of maximum rate of entropy generation at a given duration time, $\tau$, are related to the points of minimum fidelity. This is true for the other values of $\delta{\varepsilon}$ as well. In other words, as the rate of entropy generation increases, the fidelity decreases and vice versa. In a similar fashion, the regions in Fig. \ref{fig-4n}(a) where the curves for the entropy flatten or begin to flatten coincide with these changes in the entropy generation rate and the fidelity, indicating a close connection between the final entropy and fidelity of the {\footnotesize{CPHASE}} gate operation.

\begin{figure}[]
	\includegraphics[width=9.5cm]{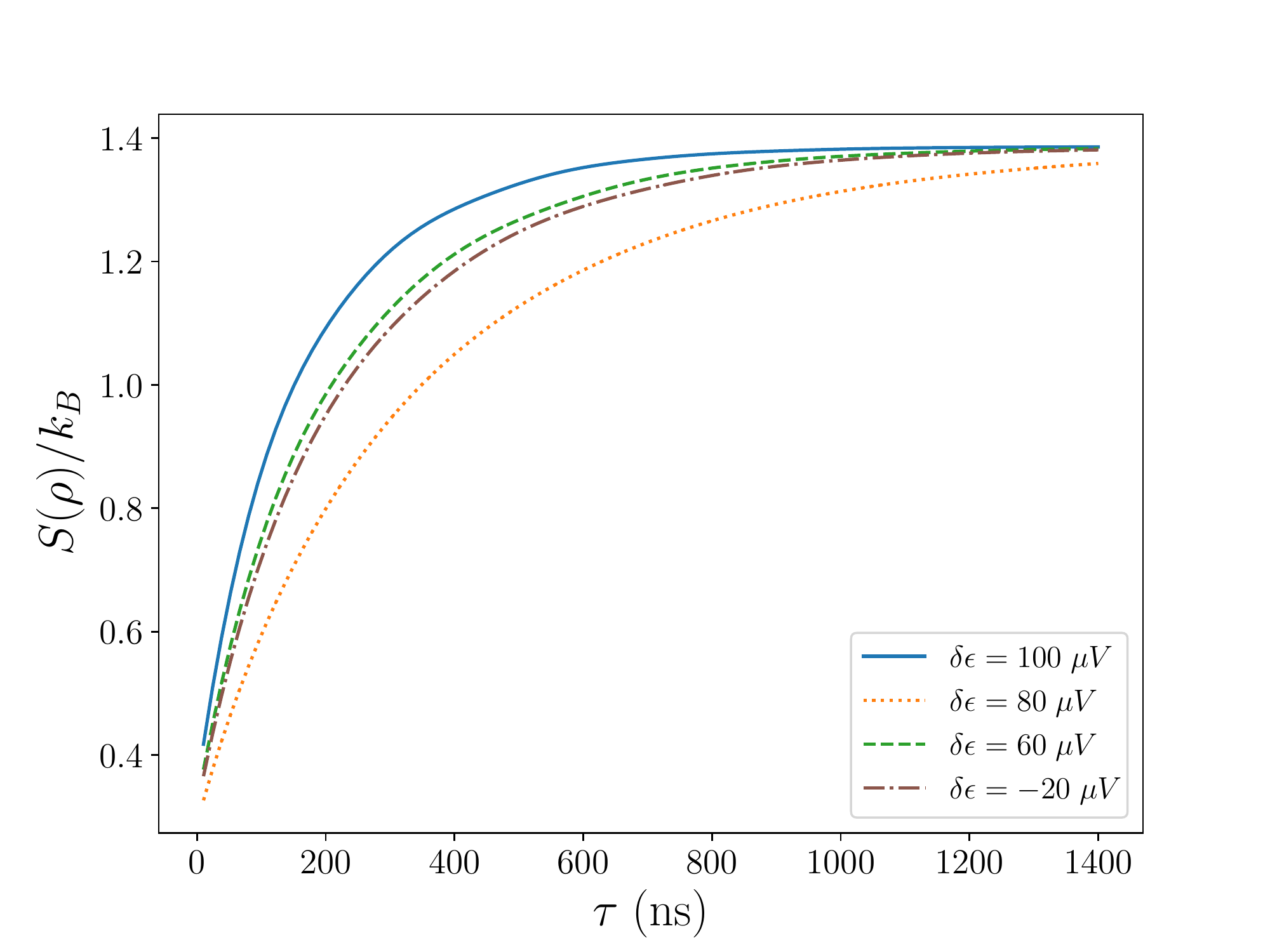} \\ (a) 
	\includegraphics[width=9.5cm]{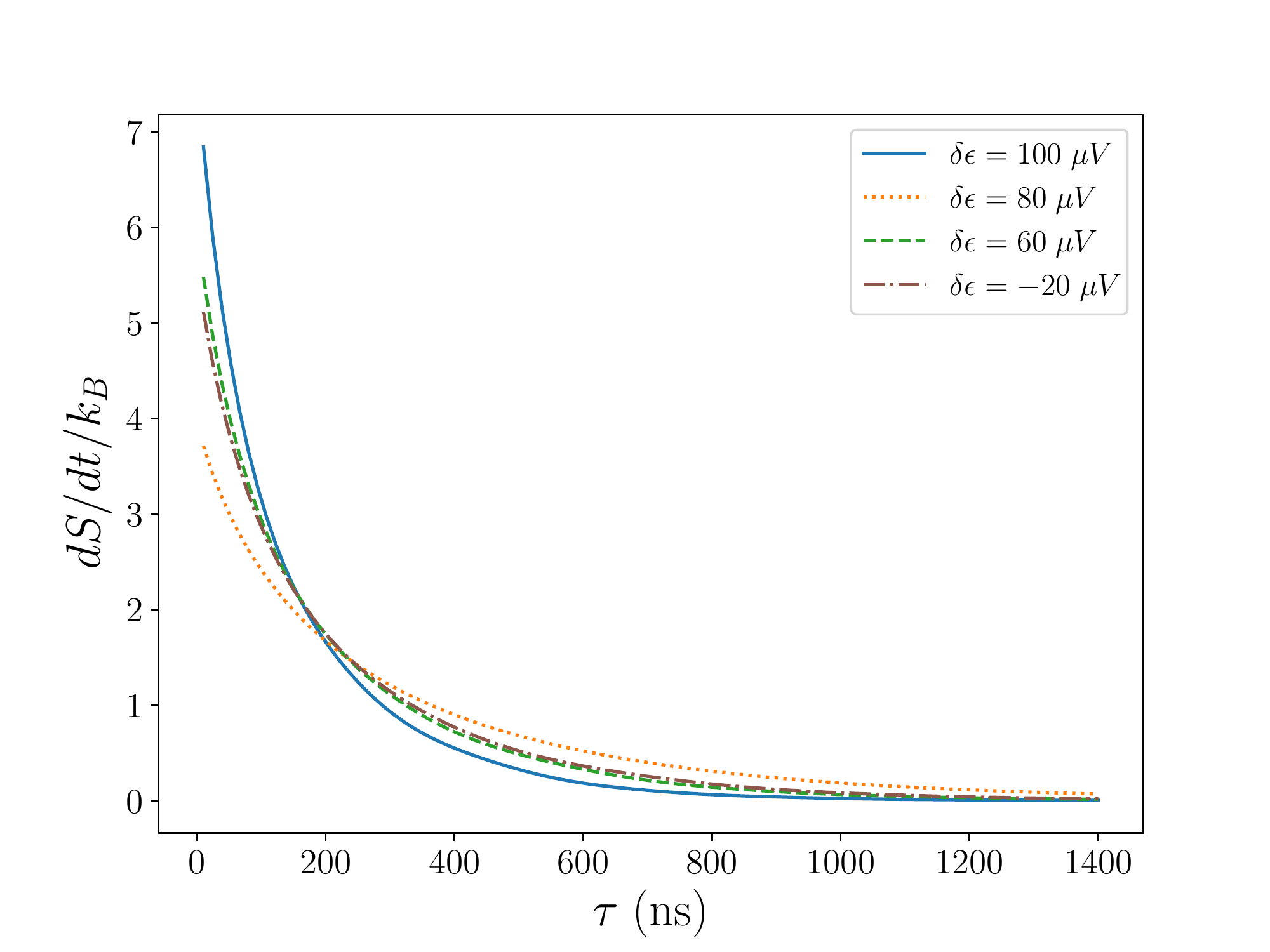} \\ (b) 
	\caption{\label{fig-4b}Final (a) entropy and (b) rate of entropy change predicted by the Lindblad equation of motion for different values of $\delta\varepsilon$ and the duration time $\tau$. These quantities are plotted in dimensionless form as $S/k_B$ and $dS/dt / k_B$.}
\end{figure}

Figure \ref{fig-4b} shows the results for the entropy and rate of entropy change \cite{EntProd} of the open system model as a function of the {\footnotesize{CPHASE}} gate duration time, $\tau$, for four different values of $\delta{\varepsilon}$. As can be seen in Fig. \ref{fig-4b}(a), for all values of $\delta{\varepsilon}$, the final entropy of the reduced system monotonically increases as a consequence of the Lindblad operators. This contrasts with the nonmonotonic increase of the final entropy predicted by the nonlinear dynamics of the SEAQT equation of motion seen in Fig. \ref{fig-4n}(a). Furthermore, with the Lindbald equation, increasing the rate of entropy change is only a consequence of the particular value of $\gamma$ used for each $\delta\varepsilon$. Figure \ref{fig-4b}(b) shows the rate of entropy change for the same four values of $\delta{\varepsilon}$. As seen, this rate monotonically decreases from a maximum value to a value that is almost zero at about $\tau = 1400 ns$. Furthermore, it is always non-negative. Note that this monotonic decrease also contrasts, as it did with the final entropy, with the results of the nonlinear dynamics of the SEAQT equation of motion seen in Fig. \ref{fig-4n}(b).

\subsection{Entanglement time prediction}

\begin{figure}[]
\includegraphics[width=9cm]{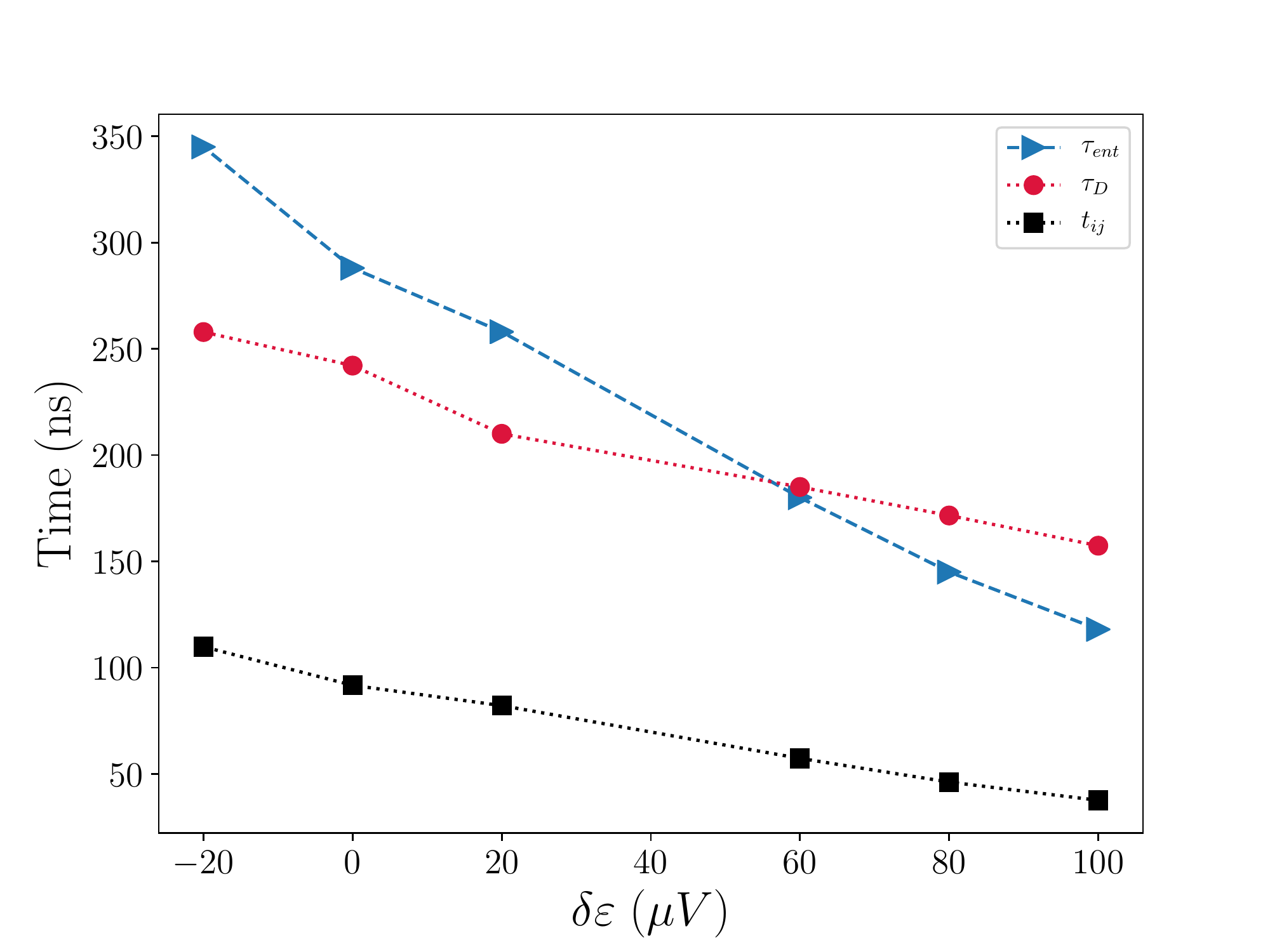}
\caption{\label{TransTime}Dissipative time $\tau_D$, maximum entanglement time $\tau_\mathrm{ent}$, and transition time $t_{ij} = 1/J_{12}$ as a function of the electrical potential of the quantum gate.}
\end{figure}

The modeling framework presented in this paper allows one to predict the fidelity of the CPHASE protocol without requiring the use of an artificial noise as is typically done (e.g., see Nichol \emph{et al.} \cite{Nichol2017}). Instead, the only free parameter required is the dissipative time that does not alter the kinetic path predicted by the SEAQT equation of motion but instead simply indicates how fast the state of the system moves along this path to either a state of stable equilibrium or some other stationary state (e.g., a Bell diagonal state). Figure \ref{TransTime} shows the relationship between the variation of the electric potential, $\delta{\varepsilon}$, and the maximum entanglement time, $\tau_\mathrm{ent}$, obtained from the experiment (blue dashed curve) in Ref.\cite{shulman2012demonstration} and the selected dissipative time, $\tau_D$, used in Eq. (\ref{Eq:Beretta}) (red dotted curve) where it is assumed that $\tau_{D_1}=\tau_{D_2}=\tau_D$. It is interesting to note that both $\tau_\mathrm{ent}$ and $\tau_D$ follow an almost linear relation with respect to the electric potential from the quantum protocol. This behavior implies that as the electric potential is increased the system evolves towards stable equilibrium faster. As is observed in Fig. \ref{fig-4n} for the case of $\delta \varepsilon = 100 \, \mu$V, the entropy evolves to stable equilibrium faster than for the case of $\delta \varepsilon = -20 \, \mu$V.

It is interesting to note that the maximum entanglement time proposed in Ref. \cite{shulman2012demonstration} produces the maximally entangled final state as a function of the electric potential. However this choice is not completely in agreement with the maximum fidelity value and, thus, produces the slight deviation from linear behavior seen in Fig. \ref{TransTime} between the electric potential and the maximum entanglement time.

\section{\label{RelDisDecay}RELATION BETWEEN THE DISSIPATIVE TIME AND THE DECAYING RATE}

Even though the dissipative time $\tau_D$ used here is based on experimental values, a theoretical expression can be developed for this particular application by relating it to the transition probabilities determined with Fermi's golden rule (e.g., \cite{Schatz1993,Chen2005}). For time-independent Hamiltonians of first order in perturbation theory, the transition probability is given by

\begin{equation}
\mathcal{P}_{ij} (t) = \frac{1}{\hbar^2} \left| \int_0^t \exp \left[ \frac{i}{\hbar} \left( E_f - E_i \right) \right] W_{fi} (t')  \right|^2,
\end{equation}
where the $E_i$ are the eigenvalues of the non-interacting Hamiltonian, $H_0$, and the $W_{fi}$ are the matrix elements of the perturbation Hamiltonian, $H'$.  Since $J_1$,$J_2 \gg J_{12}$ for the {\footnotesize{CPHASE}} gate, the only non-zero matrix element is $W_{44} = \hbar J_{12}$  and, thus, the transition probability given by Fermi's golden rule reduces to $\mathcal{P}_{ij} (t) = J_{12} t$, and the decay rate becomes

\begin{equation}
\Gamma_{ij} (t) = \frac{d}{dt}\mathcal{P}_{ij} (t).
\end{equation}

The inverse of this decay rate can then be identified with the transition time and, for the {\footnotesize{CPHASE}} gate results

\begin{equation}
t_{ij} =  \frac{1}{J_{12}},
\end{equation}
which matches in order of magnitude with the selected dissipative time. Figure \ref{TransTime} shows the linear behavior of $t_{ij}$, $\tau_D$, and $\tau_\textit{ent}$ as a function of $\delta \varepsilon$. As can be seen, $t_{ij}$ and $\tau_D$ have almost the same slope, suggesting that they are related by an average scaling factor, which in this case is approximately 3.

\section{\label{Conc}CONCLUSIONS}

In this paper, an approach based on the principle of steepest entropy ascent is used to predict the loss of entanglement or correlation that takes place during the state evolution of a CPHASE quantum gate composed of two electrons in a double quantum dot. In addition, the maximum entanglement time is shown to be proportional to the dissipative time, which in turn determines how fast the system state evolves along the unique thermodynamic path determined by the SEAQT equation of motion. The faster it moves, the greater the loss in fidelity is, i.e., the greater the final state's deviation from a Bell state is.

The effects on gate behavior of electric potential are also demonstrated, showing that increases in potential coincide with increases in the rates of entropy generation. Thus, the greater the two-qubit coupling $J_{12}$ is, the greater the rate of entropy generation and as a consequence the smaller the maximum entanglement time. These results suggest that irreversibilities as defined within the SEAQT framework are directly related to the entanglement time.
 
As pointed out in Ref. \cite{Nichol2017}, improving the maximum fidelity can be affected via the initial preparation of the qubits. The SEAQT model verifies this conclusion in good agreement with the experimental results, i.e., lower rates of entropy generation are achieved by increasing the initial fidelity of each qubit. However, the entanglement time does not depend on the initial states of the qubits but instead on the qubit coupling. For the particular case presented here, the prediction using Fermi's golden rule suggests that the entanglement time is related to the dissipation time so that, as $J_{12}$ increases, the dissipation time decreases, resulting in greater irreversibilitites. Minimizing the later would, thus, extend the entanglement time. If not possible, predicting the latter for a given coupling would provide the basis for error correction.

Finally, the results obtained suggest that the SEAQT framework is a reasonable, physically (as opposed to stochastically, e.g., Ref.  \cite{Nichol2017}) based approach for predicting the loss of entanglement of quantum computing systems. Of course, the full extent of its capabilities to model these and other related phenomena is still being explored and is the subject of future work.

\section*{ACKNOWLEDGMENTS}

This work was partially supported by the Universidad de Guanajuato under Grant No. DAIP-CIIC 233/2018. J.A.M.- B. thanks the National Council of Science and Technology (CONACyT), Mexico, for his Assistantship No. CVU-736083. S.C.-A. and C.E.D.-A. gratefully acknowledge the financial support of CONACyT, Mexico, under its SNI program. The authors also thank Oscar Loaiza-Brito from the Physics Department of the Universidad de Guanajuato for helpful discussions during the development of the Appendix.

\section*{\label{Appendix I}APPENDIX: PROOF OF THE PRESERVATION OF DENSITY OPERATOR POSITIVITY BY THE SEAQT EQUATION OF MOTION}

To demonstrate the preservation of density matrix pos- itivity by the SEAQT equation of motion, two proofs are presented here: a heuristic one and a proof based on a theorem developed by Weinberg \cite{Weinberg:2014ewa} dealing with general symmetry transformations forming groups. The latter suggests that it is sufficient and necessary, at least for density matrix transfor- mations represented by compact groups, to demonstrate that a given zero eigenvalue belonging to a particular eigenvector of the density matrix
 
\begin{equation}\label{eq:basis}  
	\rho | v \rangle = \langle v | \rho = 0
\end{equation}
remains zero after a time perturbation. This implies that, if  $\rho$ starts its time evolution with all its eigenvalues positive, they will remain non-negative at all times since anyone of these that decreases to zero will remain so from that moment on. Thus, to first order in the perturbation, $\rho$ preserves the zero eigenvalue subject to the dynamics of the SEAQT equation of motion. Now, to prove that in general a zero eigenvalue remains zero during a time evolution, it is sufficient to prove that it remains zero after a time perturbation of the density matrix for the correlated and uncorrelated cases.

For indivisible systems, the density matrix can be written, without loss of generality, in the Pauli basis as \cite{gamel2016entangled}
\begin{equation}
\rho = \frac{1}{4} a_{\mu \nu} \sigma_\mu \otimes \sigma_\nu ,
\end{equation}
where $\sigma_\mu = \{ \mathbb{I}, \sigma_1, \sigma_2, \sigma_3\}$ and the coefficients $a_{\mu \nu}$ contain all the physical information of the density matrix. In fact, the components $a_{0,j} = \vec u^T$ and $a_{j,0} =\vec v$  (for $j={1,2,3}$) are the Bloch vector components of subsystems 1 and 2, respectively, and the terms $a_{ij}$ for $i=\{1,2,3\}$ are a measure of the correlations between subsystems. For the case when there are no correlations, the elements $a_{ij}$ are all zero, and it is observed that the density matrix can be written as
\begin{equation}
\rho = \rho_1 \otimes \mathbb{I}  + \mathbb{I} \otimes \rho_2.
\end{equation}
Here, the Hilbert space is given by $\mathcal{H} = \mathcal{H}_1 \otimes \mathcal{H}_2$, and the eigenvector of Eq. (\ref{eq:basis}) is given by $|v \rangle =| v_1 \rangle \otimes | v_2 \rangle$. To first order, the density matrix for the uncorrelated case changes over time $t+ \delta t$ as
\begin{equation}
\langle v |  \rho \left( t+\delta t \right)  | v \rangle =  -\langle v |  \frac{1}{\tau_{D_1}} \tilde D_1 \otimes \rho_{2} +\frac{1}{\tau_{D_2}} \rho_1 \otimes \tilde D_2  | v \rangle \delta t 
\end{equation}
for $\tau_{D_J} \in \mathbb{R}_{+}$. For a large $\tau_{D_J}$ the perturbation vanishes and the density matrix is mapped to a positive matrix, as desired. For a non-zero perturbation, the $\tilde D_J$ operator can be written as
\begin{equation}
\tilde D_J = \sqrt{\rho_J} \left( B \log \rho \right)^J + \alpha \sqrt{\rho_J} + \beta \sqrt{\rho_J} H^J    
\end{equation}  
for $\alpha,\beta \in \mathbb{C}$. Thus, to maintain the positiveness of $\rho$ it is a sufficient condition that the term
\begin{align} \label{eq:pos}
\frac{1}{\tau_{D_1}}  \left( \rho_1 \left( B \log \rho \right)^1 + \alpha_1^* \rho_1 + \beta_1^*   H^1  \rho_1  \right) \otimes \rho_2 | v \rangle +  \\ \nonumber
\frac{1}{\tau_{D_2}} \rho_1 \otimes  \left(  \rho_2 \left( B \log \rho \right)^2 +  \alpha_2^*  \rho_2+ \beta_2^*     H^2 \rho_2 \right)  | v \rangle &=  0 
\end{align}
has a zero eigenvalue. Because of  the factorization of the Hilbert space for an uncorrelated system, the perturbation term subsystem 1 can be written as
\begin{equation}
\langle v_1 | \left( \rho_1 \left( B \log \rho \right)^1 + \alpha_1^* \rho_1+ \beta_1^*   H^1  \rho_1 \right) | v_1 \rangle \otimes \langle v_2 | \rho_2   | v_2 \rangle  =  0 ,
\end{equation}
which vanishes for $\rho_i | v_i \rangle = \langle v_i | \rho_i$ and similarly for subsystem 2. 

For the case of a correlated system, the density matrix cannot in general be factorized. Thus,  Eq. (\ref{eq:pos}) reduces to
\begin{equation}
\langle v | \rho \left( t + \delta t \right) | v \rangle = \langle v | \left( \rho \left( B \log \rho \right) + \alpha^* \rho + \beta^* \rho H \right)  | v \rangle,
\end{equation} 
which vanishes by virtue of Eq. (\ref{eq:basis}).

\begin{figure}[]
\includegraphics[width=9cm]{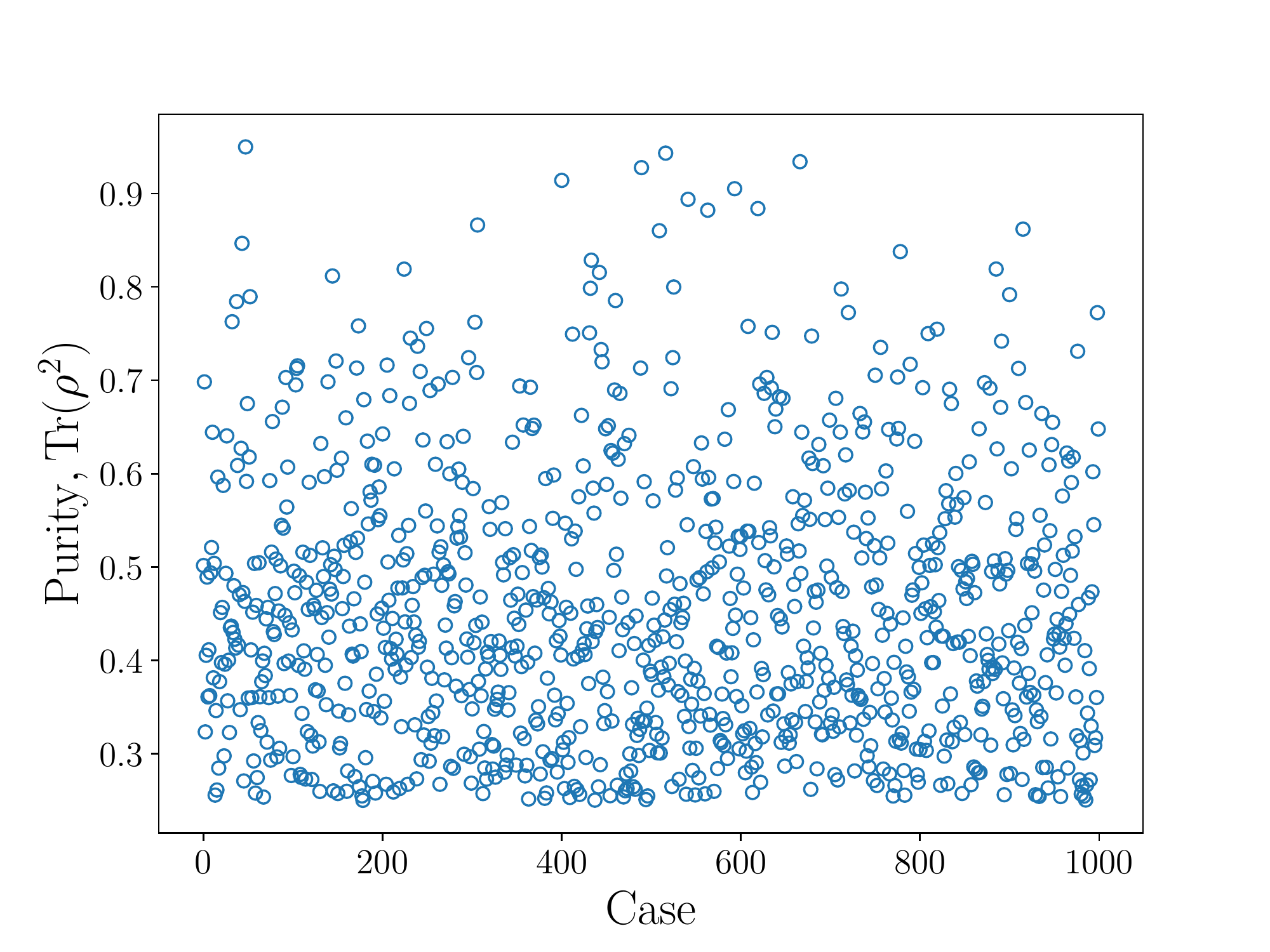}
\caption{\label{Purity}Purity for the initial condition for the 1000 random cases tested.}
\end{figure}

\begin{figure}[h!]
\includegraphics[width=9cm]{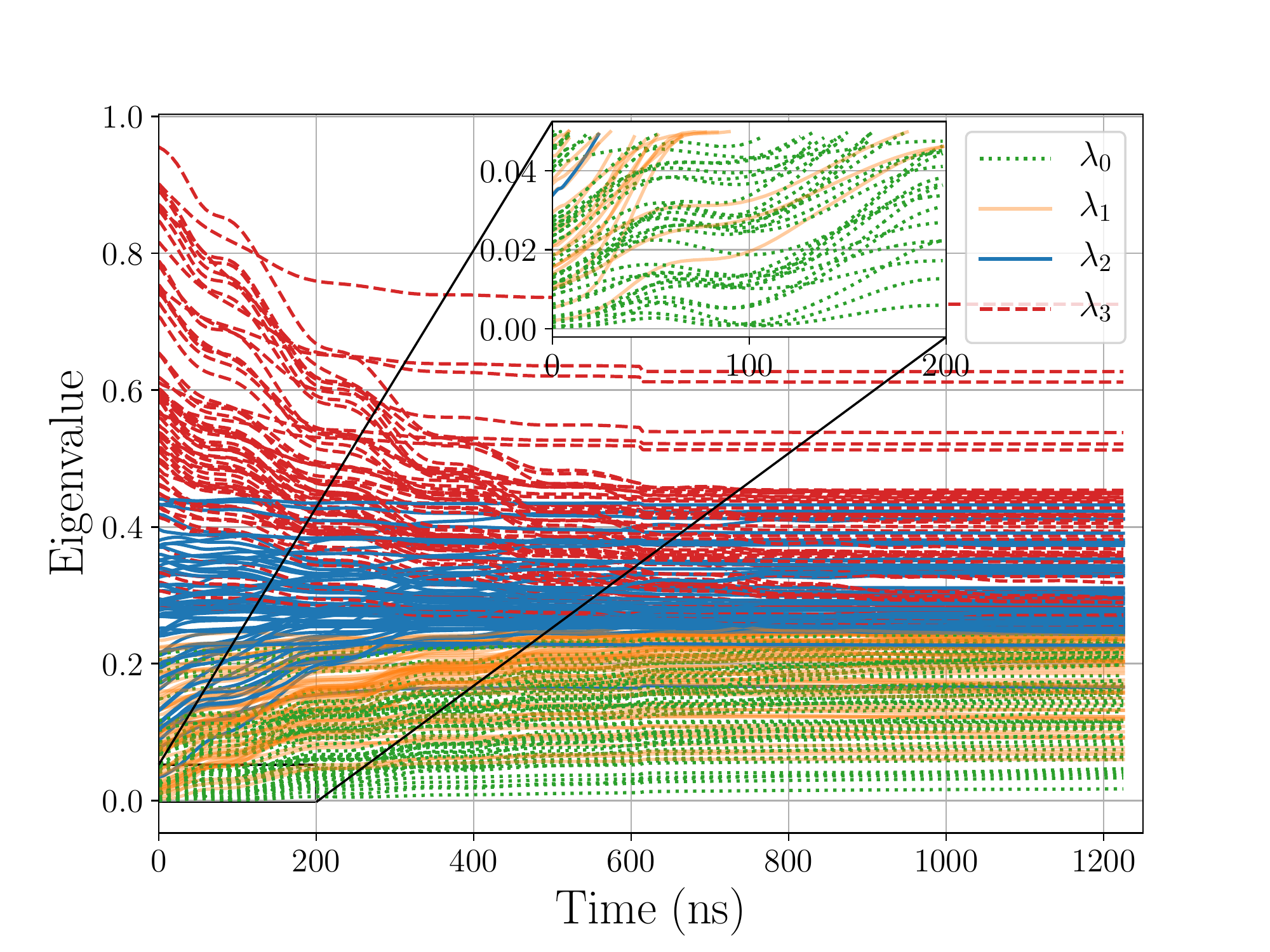}
\caption{\label{eigen0}Time evolution of the four eigenvalues of the density matrix for 62 out of the 1000 cases shown in Fig. \ref{Purity} using the SEAQT equation of motion and the Hamiltonian of Shulman \emph{et al.} \cite{shulman2012demonstration}.}
\end{figure}

\begin{figure}[h!]
	\includegraphics[width=9cm]{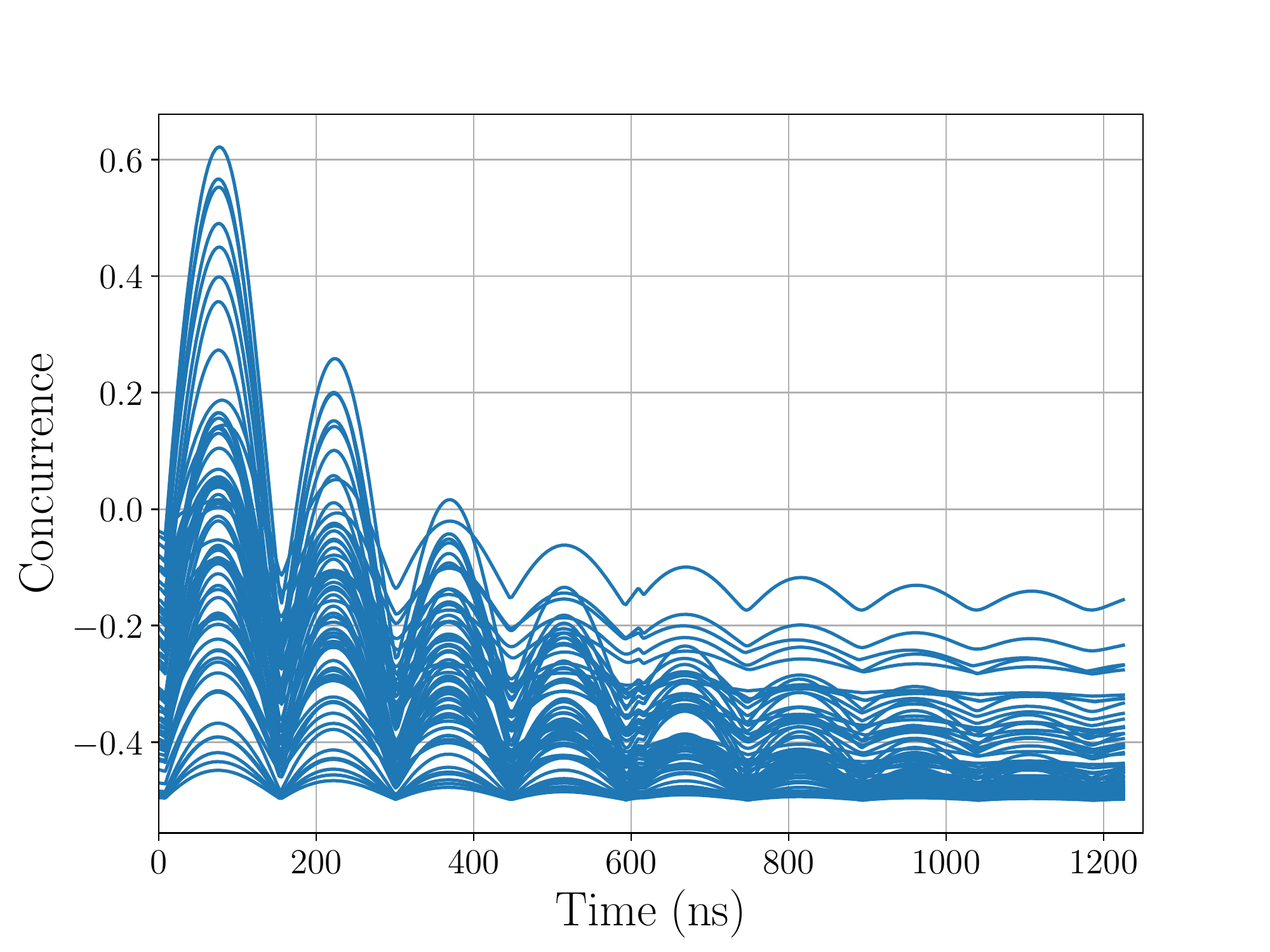}
	\caption{\label{Concurrence}Time evolution of the concurrence of the density matrix for 62 out of the 1000 cases shown in Fig. \ref{Purity} using the SEAQT equation of motion and the Hamiltonian of Shulman  \emph{et al.} \cite{shulman2012demonstration}.}
\end{figure}

As a heuristic support to the proof given above, the time evolutions of the density operator using the SEAQT equation of motion and the Hamiltonian of Shulman \emph{et al.} \cite{shulman2012demonstration} are generated for 1000 random cases of the initial state, and the evolutions of the four eigenvalues of each density matrix are observed. Figure \ref{Purity} presents the purity, $\mathrm{Tr}\, \rho^2$, of each of the 1000 initial states considered. The figure shows a wide distribution of initial conditions for the two-qubit system. Figure \ref{eigen0} shows the results for the evolutions of the four eigenvalues for each density matrix of 62 representative cases out of the 1000 tested, while Fig. \ref{Concurrence} shows their concurrence, i.e., their level of entanglement. As can be seen, all eigenvalues remain positive throughout the evolutions for these cases (as it does for the remaining 938 cases not shown here), confirming that, at least for the 1000 cases tested, the time evolution of the density matrix is always positive within the SEAQT framework.

Finally, two additional heuristic proofs are found in the literature: one in Cano-Andrade  \emph{et al.} \cite{Cano2015} and the other in Holladay \cite{Holladay2019}. In the former, several thousand initial density operators are randomly generated for a four-level correlated system with a distribution of purities similar to that seen above. In all cases, the positivity of the density operator is maintained by the SEAQT equation of motion. In the latter, entanglement evolutions of a few thousand perturbed Bell diagonal states (i.e., maximally entangled states) are presented. Two different approaches --i.e., a weighted-average perturbation approach and a general bipartite perturbation approach with a constant energy constraint and alternatively constant energy and entropy constraints-- are used to randomly generate the initial states used by the SEAQT equation of motion. In all cases, the positivity of each density operator is maintained throughout the evolutions.

\newpage

\bibliographystyle{ieeetr}

\end{document}